\documentclass[aps,prd,reprint,10pt,superscriptaddress,floatfix,nofootinbib]{revtex4-2}

\usepackage{amsmath}
\usepackage[T1]{fontenc}
\usepackage[utf8]{inputenc}
\usepackage{lmodern}
\usepackage{amssymb}
\usepackage{natbib}
\usepackage{graphicx}
\usepackage{hyperref}
\usepackage{orcidlink}
\usepackage{bm}
\usepackage{booktabs}
\usepackage{array}
\usepackage{microtype}

\graphicspath{{figures/}}

\hypersetup{colorlinks=true,linkcolor=blue,citecolor=blue,urlcolor=blue}

\newcommand{\dd}{\mathrm{d}}
\newcommand{\ii}{\mathrm{i}}
\newcommand{\ee}{\mathrm{e}}
\newcommand{\id}{\mathbb{I}}
\newcommand{\Lag}{\mathcal{L}}
\newcommand{\Oh}{\mathcal{O}}
\newcommand{\abs}[1]{\left|#1\right|}
\newcommand{\norm}[1]{\left\|#1\right\|}
\newcommand{\betaD}{\beta_{\mathrm D}}

\begin{document}
	
	\title{Dirac oscillator in a self-gravitating cosmic string:\\
		confinement spectra beyond the conical approximation}
	
	\author{Edilberto O. Silva\orcidlink{0000-0002-0297-5747}}
	\email[Edilberto O. Silva - ]{edilberto.silva@ufma.br}
	\affiliation{
		Programa de P\'os-Gradua\c c\~ao em F\'{\i}sica \&
		Coordena\c c\~ao do Curso de F\'{\i}sica -- Bacharelado,
		Universidade Federal do Maranh\~{a}o,
		65085-580 S\~{a}o Lu\'{\i}s, Maranh\~{a}o, Brazil}
	
	\date{September 8, 2026}
	
	\begin{abstract}
The gravitational field of a physical cosmic string is regular on the axis, curved across a finite vortex core, and only asymptotically conical. We test whether the bound-state spectrum of a transverse Dirac oscillator resolves this structure beyond the ideal-cone approximation. The fermion is treated as a test field on self-consistent Einstein--Abelian-Higgs vortex backgrounds, and the radial problem is cast as a generalized Hermitian eigenvalue problem. The numerical formulation is checked against the exact Minkowski and ideal-cone spectra and by independent shooting calculations. For a reference vortex family, the finite-core correction grows from the weak-confinement regime toward the percent level. When the normalized asymptotic cone is held fixed, the spectrum still varies with the measured Higgs-core radius, and the sampled backgrounds bracket a change of sign of the ground-state correction. The bound-state spectrum therefore contains information about the resolved core profiles that is not fixed by the asymptotic cone alone.

Keywords: Dirac oscillator; self-gravitating cosmic string; Einstein--Abelian--Higgs vortex; bound-state spectrum; finite-core effects.
\end{abstract}
	
	\maketitle
	
	\section{Introduction}
\label{sec:introduction}

Cosmic strings are among the most studied topological relics of symmetry-breaking phase transitions in the early Universe~\cite{NPB.1973.61.45,PRD.1985.32.1323}. In the widely used idealization, the string is an infinitely thin line whose exterior geometry is an exact cone: locally flat, with all of the physics encoded in a single angular deficit $\delta=2\pi(1-b)$. A physical string produced by a self-gravitating vortex is qualitatively richer. Solving the Einstein--Abelian-Higgs equations yields a metric that is regular on the axis, curved throughout a finite transition region set by the core width, and conical only asymptotically~\cite{PRD.1985.32.1323,PRD.1999.60.125012,PRD.2000.62.085004,CQG.2015.32.155001}. The question we address is whether a confined relativistic quantum system, as opposed to a scattering or asymptotic observable, can resolve this difference.

The Dirac oscillator is one of the very few exactly solvable models of relativistic confinement. The idea of appending an oscillator-like term to the Dirac equation can be traced to early studies of relativistic wave equations~\cite{ZP.1930.62.677,PR.1969.180.1225,PTP.1973.49.2158}, but the model acquired its modern form and name with Moshinsky and Szczepaniak~\cite{JPA.1989.22.L817}, whose nonminimal substitution $\bm p\rightarrow\bm p-\ii M\omega\betaD\bm r$ yields a linear-in-momentum, exactly solvable spectrum with a strong spin--orbit coupling and a clean nonrelativistic harmonic-oscillator limit. Its algebraic structure was quickly mapped out: a hidden supersymmetry~\cite{PRL.1990.64.1643}, an underlying symmetry Lie algebra and ladder operators~\cite{JPA.1990.23.2263,JPA.1991.24.667}, and a transparent link to spin--orbit dynamics~\cite{PLA.1991.158.19}. The planar two-dimensional reduction relevant to layered and defect systems was solved exactly by Villalba~\cite{PRA.1994.49.586}, further examined in the $(2+1)$-dimensional setting~\cite{EPL.2014.108.30003}, and later generalized to arbitrary spin~\cite{JPA.1996.29.4217}. A close correspondence with quantum optics, the Jaynes--Cummings and anti-Jaynes--Cummings models, was subsequently established~\cite{JPA.1999.32.5367,PRA.2007.76.041801}, and the model was explored under Aharonov--Bohm fluxes~\cite{PLA.2004.325.21}, external magnetic fields~\cite{PLA.2010.374.1021}, and Coulomb-type tensor couplings~\cite{JPA.2007.40.6427}. Extensive work has since characterized its deformation under noncommutative geometry~\cite{CTP.2004.42.664,IJTP.2010.49.1699,PRA.2014.90.042111}, a minimal length or generalized uncertainty principle~\cite{JPA.2005.38.1747,JPA.2006.39.10909,JMP.2007.48.113508,PRD.2015.91.045032,JMP.2017.58.063504,EPL.2019.128.30004}, and a $\kappa$-deformation of the underlying algebra~\cite{PLB.2014.731.327,PLB.2014.738.44}, as well as its thermodynamic and statistical properties~\cite{PLA.2003.311.93,EPJP.2013.128.124,EPL.2014.108.10005,FBS.2015.56.115,JPA.2020.53.185204,PA.2020.553.124207}. The model also exhibits collective phenomena such as a chirality quantum phase transition~\cite{PRA.2008.77.063815}. Photonic and microwave realizations were reported in Refs.~\cite{OL.2010.35.1302,PRL.2013.111.170405}, with a review in Ref.~\cite{JPA.2017.50.081001}. Related formulations have since been explored in graphene and Dirac materials~\cite{PS.2015.90.045702,PLA.2016.380.773}, nuclear-structure bases~\cite{PRC.2020.102.054308}, quantum measurement~\cite{PRA.2021.104.022602}, and information-theoretic analyses~\cite{PRA.2023.108.022812}.

A parallel and, for our purpose, more directly relevant line of work has placed the Dirac oscillator in curved and topologically nontrivial backgrounds. Starting from the oscillator in the field of a topological defect~\cite{PRA.2011.84.032109}, later studies incorporated rotation~\cite{GRG.2013.45.1847}, spin and magnetic flux in the magnetic cosmic-string spacetime~\cite{EPJC.2014.74.3187}, vector and scalar potentials with spin and pseudospin symmetries~\cite{EPJC.2019.79.596}, generalized oscillator couplings~\cite{AHEP.2018.2018.2741694}, gravity's-rainbow and Lorentz-violating deformations~\cite{EPJP.2018.133.409}, spinning strings with torsion~\cite{EPJC.2019.79.311,PRD.2020.102.105020}, and fully curved-spacetime formulations~\cite{PS.2020.95.055304}. Recent work has considered global-monopole backgrounds~\cite{PS.2025.100.095308}, clouds of strings~\cite{EPJC.2026.86.506}, spinning cosmic strings~\cite{NPB.2026.1029.117506}, and non-Abelian extensions of the oscillator itself~\cite{MPLA.2026.41.2650143}. In these studies, however, the string is represented by an idealized conical, or otherwise singular, metric: it is taken to be infinitely thin, and its finite, self-gravitating core plays no role.

The Dirac oscillator is particularly suitable for this question because it carries a tunable confinement length of order $(M\omega)^{-1/2}$, which lets it probe the geometry at a chosen scale: a weakly confined state spreads into the asymptotic cone, whereas a strongly confined state is squeezed into the regular core. While scalar and spinor fields have been analyzed in genuine gravitating-string geometries~\cite{CQG.2021.38.205006,CQG.2025.42.225017}, a confined bound-state calculation performed as a test-field problem directly on a self-consistent gravitating-vortex background, and compared level by level against the conical idealization, has to our knowledge been missing.

The paper is organized as follows. Sections~\ref{sec:geometry} and \ref{sec:radial} formulate the covariant Dirac oscillator in a generic regular string background, derive the coupled radial system, and expose its generalized Hermitian structure. Section~\ref{sec:results} presents the exact Minkowski and ideal-cone benchmarks that anchor the numerical formulation. Section~\ref{sec:vortex} summarizes the self-gravitating Einstein--Abelian--Higgs vortex backgrounds used in the spectral calculation, while extended numerical diagnostics for both the spectral solver and the background solver are collected in the Supplemental Material. Section~\ref{sec:spectra} contains the physical spectral comparison between the resolved vortex and the corresponding ideal cone. The central result is that the finite gravitating core leaves a distinct spectral imprint that the asymptotic pair $(\bar a,b)$ alone does not determine.

To our knowledge, this is the first bound-state study of a Dirac oscillator treated as a test field on a self-consistent, self-gravitating cosmic-string background, and the first demonstration within this model that the confinement spectrum can resolve the finite core beyond the asymptotic deficit. In this sense, the calculation turns a confined quantum probe into a spectroscopic test of string microphysics: two strings with identical normalized far-field geometry but different internal profiles can, in principle, be distinguished by their confinement spectra, even though that distinction is not encoded in the deficit angle alone. The same mechanism may also be relevant to engineered Dirac systems in which regularized defect geometries can be realized on experimentally accessible length scales.

\section{Gravitating-string geometry}
	\label{sec:geometry}
	
	\subsection{Line element and regularity conditions}
	
	We consider the static, cylindrically symmetric, and boost-invariant line element
	\begin{equation}
		\dd s^2=N^2(r)\dd t^2-\dd r^2-L^2(r)\dd\varphi^2-N^2(r)\dd z^2,
		\label{eq:metric}
	\end{equation}
	with $r\geq0$ and $0\leq\varphi<2\pi$.  Here $N(r)>0$ is the dimensionless metric lapse that multiplies the temporal and axial directions, and $L(r)\geq0$ is the circumferential radius, with the same dimension as the proper radial coordinate $r$, so that a coordinate circle has proper length $2\pi L(r)$.  In the dimensionless vortex variables introduced in Sec.~\ref{sec:vortex}, the numerical coordinate is scaled with $(e\eta)^{-1}$: $r=e\eta\,r_{\rm phys}$ and $L=e\eta\,L_{\rm phys}$.  Since $m_{\rm v}=\sqrt{2}\,e\eta$, one dimensionless radial unit corresponds to $(e\eta)^{-1}=\sqrt{2}\,m_{\rm v}^{-1}$, not to $m_{\rm v}^{-1}$ itself.  Their axis and asymptotic values are fixed in Eqs.~\eqref{eq:regular-axis}--\eqref{eq:asymptotic}.  Boost invariance along $z$ forces the $g_{tt}$ and $g_{zz}$ factors to coincide, which is why a single function $N(r)$ multiplies both $\dd t^2$ and $\dd z^2$.  The metric determinant and the invariant volume element are
	\begin{equation}
		g=-N^4L^2,
		\qquad
		\sqrt{-g}=N^2L.
		\label{eq:det}
	\end{equation}
	Because $g_{rr}=-1$, the coordinate $r$ measures proper radial distance, a property we use repeatedly below.
	
	A regular resolved axis requires that the circumferential radius $L$ vanish linearly and that the lapse be smooth and extremal there,
	\begin{equation}
		L(0)=0,
		\quad L'(0)=1,
		\quad N(0)=N_0>0,
		\quad N'(0)=0.
		\label{eq:regular-axis}
	\end{equation}
	The condition $L'(0)=1$ guarantees the absence of a conical singularity at $r=0$: a small circle of proper radius $r$ has circumference $2\pi L(r)\to2\pi r$, so the axis looks locally like flat two-dimensional space.  In the exterior the profiles approach
	\begin{equation}
		N(r)\longrightarrow a,
		\qquad
		L(r)\longrightarrow br+c,
		\qquad 0<b\leq1.
		\label{eq:asymptotic}
	\end{equation}
	The deficit angle is $\delta=2\pi(1-b)$.  In the exactly conical exterior, the additive constant $c$ can be removed locally by the radial translation $\rho=r+c/b$, after which $L=b\rho$ and $\dd\rho=\dd r$; thus $c$ is not an additional invariant of the asymptotic cone.  Globally, however, the resolved geometry has a distinguished regular axis at $r=0$, so the full profiles $N(r)$ and $L(r)$ through the core, rather than the constants $a$, $b$, and $c$ alone, determine the finite-core spectrum.
	
	The geometry is generated by an Einstein--vortex sector.  Schematically,
	\begin{equation}
		S_{\mathrm{bg}}=\int\dd^4x\sqrt{-g}\left[\frac{R}{16\pi G}+\Lag_{\mathrm{vortex}}\right].
		\label{eq:bg-action}
	\end{equation}
	The fermion studied below is a test field and is not included in Eq.~\eqref{eq:bg-action}.  Gravity therefore enters its equation only through the tetrad and spin connection, not through an additive $R/(16\pi G)$ term in the first-order Dirac operator.
	
	\subsection{Orthonormal coframe and spin connection}
	\label{sec:coframe}
	
	A convenient diagonal orthonormal coframe $\vartheta^{\hat a}$, for which $\dd s^2=\eta_{\hat a\hat b}\vartheta^{\hat a}\vartheta^{\hat b}$ with $\eta=\mathrm{diag}(+,-,-,-)$, is
	\begin{equation}
		\vartheta^{\hat0}=N\dd t,
		\quad
		\vartheta^{\hat1}=\dd r,
		\quad
		\vartheta^{\hat2}=L\dd\varphi,
		\quad
		\vartheta^{\hat3}=N\dd z.
		\label{eq:coframe}
	\end{equation}
	The Levi-Civita connection one-forms $\omega^{\hat a}{}_{\hat b}$ follow from Cartan's first structure equation with vanishing torsion,
	\begin{equation}
		\dd\vartheta^{\hat a}+\omega^{\hat a}{}_{\hat b}\wedge\vartheta^{\hat b}=0,
		\qquad
		\omega_{\hat a\hat b}=-\omega_{\hat b\hat a}.
		\label{eq:cartan}
	\end{equation}
	Taking exterior derivatives of Eq.~\eqref{eq:coframe} and using $\dd r=\vartheta^{\hat1}$, $\dd t=\vartheta^{\hat0}/N$, $\dd\varphi=\vartheta^{\hat2}/L$, and $\dd z=\vartheta^{\hat3}/N$ gives
	\begin{align}
		\dd\vartheta^{\hat0}&=N'\dd r\wedge\dd t=-\frac{N'}{N}\vartheta^{\hat0}\wedge\vartheta^{\hat1},
		\label{eq:dtheta0}\\
		\dd\vartheta^{\hat2}&=L'\dd r\wedge\dd\varphi=-\frac{L'}{L}\vartheta^{\hat2}\wedge\vartheta^{\hat1},
		\label{eq:dtheta2}\\
		\dd\vartheta^{\hat3}&=N'\dd r\wedge\dd z=-\frac{N'}{N}\vartheta^{\hat3}\wedge\vartheta^{\hat1},
		\label{eq:dtheta3}
	\end{align}
	while $\dd\vartheta^{\hat1}=0$.  Matching these against Eq.~\eqref{eq:cartan} identifies the only nonvanishing connection forms,
	\begin{equation}
		\omega^{\hat0}{}_{\hat1}=\frac{N'}{N}\vartheta^{\hat0},
		\quad
		\omega^{\hat2}{}_{\hat1}=\frac{L'}{L}\vartheta^{\hat2},
		\quad
		\omega^{\hat3}{}_{\hat1}=\frac{N'}{N}\vartheta^{\hat3}.
		\label{eq:connection-forms}
	\end{equation}
	Each is proportional to the corresponding coframe leg contracted with the radial direction.
	
	\subsection{Contracted Dirac operator and the radial connection}
	\label{sec:diracop}
	
	The spinor covariant derivative is $\nabla_\mu=\partial_\mu+\Gamma_\mu$ with
	\begin{equation}
		\Gamma_\mu=\tfrac18\,\omega_{\hat a\hat b\mu}\,[\gamma^{\hat a},\gamma^{\hat b}],
		\label{eq:Gamma}
	\end{equation}
	where $\omega_{\hat a\hat b\mu}$ are the components of Eq.~\eqref{eq:connection-forms}.  Contracting directly with $\gamma^\mu=E_{\hat a}{}^\mu\gamma^{\hat a}$, where $E_{\hat a}{}^\mu$ is the inverse tetrad, the connection pieces collect into a single radial term.  Using the components in Eq.~\eqref{eq:connection-forms}, one obtains
	\begin{equation}
		\gamma^\mu\Gamma_\mu=\tfrac12\,\gamma^{\hat1}\,\frac{\dd}{\dd r}\ln\!\left(N^2L\right)
		=\gamma^{\hat1}A(r),
		\label{eq:contracted}
	\end{equation}
	so that the full contracted operator becomes
	\begin{equation}
		\ii\gamma^\mu\nabla_\mu=
		\frac{\ii\gamma^{\hat0}}{N}\partial_t
		+\ii\gamma^{\hat1}\left(\partial_r+A\right)
		+\frac{\ii\gamma^{\hat2}}{L}\partial_\varphi
		+\frac{\ii\gamma^{\hat3}}{N}\partial_z,
		\label{eq:dirac-operator}
	\end{equation}
	with
	\begin{equation}
		A(r)=\frac{N'}{N}+\frac{L'}{2L}
		=\frac{\dd}{\dd r}\ln\!\left(N\sqrt{L}\right)
		=\tfrac12\,\frac{\dd}{\dd r}\ln\!\left(N^2L\right).
		\label{eq:A}
	\end{equation}
	Equation~\eqref{eq:A} shows that $A=(\ln N\sqrt L)'$, the amplitude factor removed by the field redefinition in Sec.~\ref{sec:radial}.  For $N=1$ and $L=r$, Eq.~\eqref{eq:dirac-operator} reduces to the standard cylindrical form with $A=1/(2r)$.
	
	\section{Covariant Dirac oscillator and radial reduction}
	\label{sec:radial}
	
	\subsection{Nonminimal substitution and separation}
	
	The radial Dirac oscillator is defined in the local Lorentz frame by the anti-Hermitian momentum shift
	\begin{equation}
		p_{\hat r}\longrightarrow p_{\hat r}-\ii M\omega\betaD r,
		\qquad \betaD=\gamma^{\hat0}.
		\label{eq:substitution}
	\end{equation}
	Because $g_{rr}=-1$, the coordinate $r$ appearing here is the proper radial distance, so no additional metric factor is needed in the oscillator coupling.  The field equation is
	\begin{equation}
		\left[\ii\gamma^\mu\nabla_\mu
		+\ii M\omega r\,\gamma^{\hat1}\betaD-M\right]\Psi=0.
		\label{eq:covariant-do}
	\end{equation}
	We set the longitudinal momentum to zero ($k_z=0$, the transverse sector) and separate variables with the ansatz
	\begin{equation}
		\Psi_{js}=\ee^{-\ii Et+\ii j\varphi}
		\begin{pmatrix}
			f(r)\chi_s\\
			-\ii v(r)\chi_{-s}
		\end{pmatrix},
		\qquad
		\sigma_3\chi_s=s\chi_s,
		\label{eq:ansatz}
	\end{equation}
	where $s=\pm1$ labels the $\sigma_3$ eigenvalue of the upper two-spinor, while the lower two-spinor belongs to the opposite sector $\chi_{-s}$; $j\in\mathbb Z+\tfrac12$ is the total angular momentum.
	
	\subsection{Block reduction to the first-order radial system}
	\label{sec:block}
	
	In the Dirac representation, Eq.~\eqref{eq:covariant-do} with $k_z=0$ reads
	\begin{equation}
		\left[\frac{E}{N}\gamma^{\hat0}
		+\ii\gamma^{\hat1}\left(D_r+M\omega r\betaD\right)
		+\frac{\ii\gamma^{\hat2}}{L}\partial_\varphi-M\right]\Psi=0,
		\label{eq:app-dirac}
	\end{equation}
	where $D_r=\partial_r+A$.  Writing $\Psi=(\phi,\chi)^{T}$ in terms of two 2-spinors and using $\gamma^{\hat0}=\mathrm{diag}(\id_2,-\id_2)$ and $\gamma^{\hat k}=\bigl(\begin{smallmatrix}0&\sigma_k\\-\sigma_k&0\end{smallmatrix}\bigr)$, the upper and lower blocks separate into
	\begin{align}
		\left(\frac EN-M\right)\phi+
		\left[\ii\sigma_1\!\left(D_r-M\omega r\right)
		+\frac{\ii\sigma_2}{L}\partial_\varphi\right]\chi&=0,
		\label{eq:app-upper}\\
		\left[-\ii\sigma_1\!\left(D_r+M\omega r\right)
		-\frac{\ii\sigma_2}{L}\partial_\varphi\right]\phi
		-\left(\frac EN+M\right)\chi&=0.
		\label{eq:app-lower}
	\end{align}
	Inserting $\phi=f(r)\chi_s\ee^{\ii j\varphi}$ and $\chi=-\ii v(r)\chi_{-s}\ee^{\ii j\varphi}$, and using $\sigma_1\chi_{-s}=\chi_s$, $\sigma_2\chi_{-s}=-\ii s\,\chi_s$, and $\partial_\varphi\to\ii j$, the two blocks collapse to the exact first-order radial system
	\begin{align}
		v'+\left(A-M\omega r+\frac{sj}{L}\right)v
		+\left(\frac{E}{N}-M\right)f&=0,
		\label{eq:fv1}\\
		f'+\left(A+M\omega r-\frac{sj}{L}\right)f
		-\left(\frac{E}{N}+M\right)v&=0.
		\label{eq:fv2}
	\end{align}
	The label $s$, defined through the $\sigma_3$ eigenvalue of the upper two-spinor, enters only through the combination $sj/L$, so a change $s\to-s$ at fixed $j$ is equivalent to $j\to-j$.
	
	\subsection{Removal of the connection term}
	
	The amplitude factor $A$ is eliminated by the rescaling anticipated below Eq.~\eqref{eq:A},
	\begin{equation}
		U=N\sqrt L\,f,
		\qquad
		V=N\sqrt L\,v.
		\label{eq:rescale}
	\end{equation}
	Since $A=(\ln N\sqrt L)'$, one has $f'=[U'-AU]/(N\sqrt L)$ and $v'=[V'-AV]/(N\sqrt L)$, so every explicit $A$ cancels when Eqs.~\eqref{eq:fv1}--\eqref{eq:fv2} are multiplied by $N\sqrt L$.  Defining
	\begin{equation}
		B(r)=M\omega r-\frac{sj}{L(r)},
		\qquad
		\epsilon_\pm(r;E)=\frac{E}{N(r)}\pm M,
		\label{eq:B}
	\end{equation}
	the system takes the manifestly symmetric form
	\begin{align}
		U'+BU-\epsilon_+V&=0,
		\label{eq:UV1}\\
		V'-BV+\epsilon_-U&=0.
		\label{eq:UV2}
	\end{align}
	The geometry enters through $B(r)$ and the local energy functions $\epsilon_\pm(r;E)$.  The apparent $1/r$ behavior of $B$ at a regular axis is the standard cylindrical spin--angular term and is handled by the Frobenius conditions derived below.
	
	\subsection{Generalized Hermitian eigenvalue problem}
	
	Because $\epsilon_\pm=E/N\pm M$ are affine in $E$, Eqs.~\eqref{eq:UV1}--\eqref{eq:UV2} are linear in the eigenvalue and can be written as the generalized eigenproblem
	\begin{equation}
		\mathsf H_{js}\bm Y=E\,\mathsf W\bm Y,
		\qquad
		\bm Y=\begin{pmatrix}U\\V\end{pmatrix},
		\label{eq:gevp}
	\end{equation}
	with
	\begin{equation}
		\mathsf H_{js}=
		\begin{pmatrix}
			M&-\partial_r+B\\
			\partial_r+B&-M
		\end{pmatrix},
		\qquad
		\mathsf W=\frac1N\,\id_2.
		\label{eq:HW}
	\end{equation}
	Equations~\eqref{eq:UV1}--\eqref{eq:UV2} are recovered from the two rows of Eq.~\eqref{eq:gevp}.  For regular and decaying states $\mathsf H_{js}$ is formally symmetric under the ordinary $\dd r$ product, while $\mathsf W$ is positive because $N>0$.
	
	\subsection{Formal symmetry, orthogonality, and the conserved norm}
	\label{sec:hermiticity}
	
	The only nontrivial part of the formal-symmetry claim is that the off-diagonal first-order operators are mutual adjoints.  For two radial spinors $\bm X=(X_1,X_2)^{T}$ and $\bm Y=(Y_1,Y_2)^{T}$ obeying the same regular and decaying boundary conditions, integration by parts gives
	\begin{align}
		\int_0^\infty\!\dd r\,X_1^{*}\left(-\partial_r+B\right)Y_2
		&=\int_0^\infty\!\dd r\,\left[\left(\partial_r+B\right)X_1\right]^{*}Y_2
		\notag\\
		&\quad-\bigl[X_1^{*}Y_2\bigr]_0^\infty .
		\label{eq:adjoint}
	\end{align}
	The boundary term vanishes: at $r=0$ the Frobenius powers of Sec.~\ref{sec:frobenius} make $X_1^{*}Y_2\sim r^{\alpha_U+\alpha_V+1}\to0$, and at $r\to\infty$ the Gaussian tail~\eqref{eq:gaussian} kills it.  Hence $(\partial_r+B)^\dagger=-\partial_r+B$, and $\mathsf H_{js}$ is formally symmetric on the regular, decaying domain used below.  We do not require a separate deficiency-index classification for the numerical construction: after Galerkin projection the problem is a Hermitian generalized eigenvalue problem with positive weight, so the finite-dimensional eigenvalues are real.  For two continuum eigenpairs $(E_m,\bm Y_m)$ and $(E_n,\bm Y_n)$ satisfying the same boundary conditions, contracting Eq.~\eqref{eq:gevp} and subtracting yields the weighted orthogonality relation
	\begin{equation}
		(E_m-E_n)\int_0^\infty\!\dd r\,\bm Y_n^{\dagger}\mathsf W\bm Y_m=0.
		\label{eq:weighted-orthogonality}
	\end{equation}
	The physical inner product is therefore the weighted one, and the conserved norm per unit length of the string is
	\begin{equation}
		\norm{\Psi}^2=2\pi\int_0^\infty\frac{\dd r}{N(r)}
		\left(\abs{U}^2+\abs{V}^2\right).
		\label{eq:norm}
	\end{equation}
	The factor $1/N$ is the same weight that appears in $\mathsf W$, which is why the discrete weight matrix must be kept positive definite in the numerics.
	
	\subsection{Decoupled second-order equations}
	
	For the asymptotic analysis it is convenient to eliminate one component.  Solving Eq.~\eqref{eq:UV1} for $V=(U'+BU)/\epsilon_+$ and substituting into Eq.~\eqref{eq:UV2} gives the second-order equation for the upper component,
	\begin{equation}
		U''-\frac{\epsilon_+'}{\epsilon_+}U'
		+\left[\epsilon_+\epsilon_-+B'-B^2
		-\frac{\epsilon_+'}{\epsilon_+}B\right]U=0,
		\label{eq:upper-decoupled}
	\end{equation}
	while the symmetric elimination of $U$ yields
	\begin{equation}
		V''-\frac{\epsilon_-'}{\epsilon_-}V'
		+\left[\epsilon_+\epsilon_--B'-B^2
		+\frac{\epsilon_-'}{\epsilon_-}B\right]V=0.
		\label{eq:lower-decoupled}
	\end{equation}
	The first-derivative terms carry the coefficient $\epsilon_\pm'/\epsilon_\pm$, which is singular wherever $\epsilon_+$ or $\epsilon_-$ passes through zero.  Away from such points, Eqs.~\eqref{eq:upper-decoupled}--\eqref{eq:lower-decoupled} and the coupled system~\eqref{eq:UV1}--\eqref{eq:UV2} are mathematically equivalent: one reconstructs $V=(U'+BU)/\epsilon_+$ from a solution of Eq.~\eqref{eq:upper-decoupled}, or $U=-(V'-BV)/\epsilon_-$ from Eq.~\eqref{eq:lower-decoupled}.  Consequently, solving the coupled first-order equations does not select a particular radial solution.  In the numerical analysis below we nevertheless solve the second-order equations directly as an additional check.  Equation~\eqref{eq:upper-decoupled} is used for the positive levels considered here, while Eq.~\eqref{eq:lower-decoupled} is tested in benchmark channels for which $\epsilon_-$ has no zero; whenever a denominator vanishes, the nonsingular first-order system is retained across that point.
	
	\section{Boundary conditions and exact benchmarks}
	\label{sec:benchmarks}
	
	\subsection{Frobenius analysis at the axis}
	\label{sec:frobenius}
	
	Let the circumferential radius behave near the inner endpoint as
	\begin{equation}
		L(r)=\ell r+\Oh(r^3),
		\label{eq:axis-slope}
	\end{equation}
	so that $\ell=L'(0)$.  For a resolved core $\ell=1$; for an ideal cone $\ell=b$.  Near $r=0$ the potential is dominated by its centrifugal piece, $B\simeq-sj/(\ell r)$, and $\epsilon_\pm$ tend to the finite constants $E/N_0\pm M$ with $\epsilon_\pm'\to0$.  Substituting $U\sim r^{p}$ into Eq.~\eqref{eq:upper-decoupled} and keeping the $r^{-2}$ terms $B'\simeq sj/(\ell r^2)$ and $B^2\simeq (j/\ell)^2/r^2$ gives the indicial equation
	\begin{equation}
		p(p-1)=\left(\frac{j}{\ell}\right)^{2}-s\,\frac{j}{\ell},
		\label{eq:indicial}
	\end{equation}
	whose roots are $p_\pm=\tfrac12\pm\abs{j/\ell-s/2}$.  Retaining the regular (larger) root, and repeating the argument for $V$ with the sign of $B'$ reversed in Eq.~\eqref{eq:lower-decoupled}, the admissible behaviors are
	\begin{equation}
		U\sim r^{\alpha_U+1/2},
		\qquad
		V\sim r^{\alpha_V+1/2},
		\label{eq:frobenius}
	\end{equation}
	with the slope-dependent indices
	\begin{equation}
		\alpha_U=\abs{\frac{j}{\ell}-\frac{s}{2}},
		\qquad
		\alpha_V=\abs{\frac{j}{\ell}+\frac{s}{2}}.
		\label{eq:alphas}
	\end{equation}
	The explicit appearance of $\ell$ is essential: on an ideal cone the regular basis must use $\ell=b$, rather than the resolved-core value $\ell=1$, in order to represent the correct near-axis power.
	
	\subsection{Large-radius Gaussian tail}
	
	At large radius the oscillator term dominates, $B\simeq M\omega r$.  Keeping the leading terms in either decoupled equation gives $U''-M^2\omega^2r^2U\simeq0$ and $V''-M^2\omega^2r^2V\simeq0$.  Selecting the decaying branch, both components have the Gaussian tail
	\begin{equation}
		U,V\propto\exp\!\left(-\frac{M\omega r^2}{2}\right),
		\label{eq:gaussian}
	\end{equation}
	which is the decaying behavior imposed at the outer end of the integration domain and the tail built into the Laguerre basis of Sec.~\ref{sec:galerkin}.
	
	\subsection{Minkowski space}
	\label{sec:flat}
	
	For $N=1$ and $L=r$ we have $\epsilon_\pm=E\pm M$ (constant), $\epsilon_\pm'=0$, and $B=M\omega r-sj/r$.  Define $m_{js}=j-s/2\in\mathbb Z$.  Then $B'-B^2=M\omega+sj/r^2-M^2\omega^2r^2+2M\omega sj-j^2/r^2$, and Eq.~\eqref{eq:upper-decoupled} becomes
	\begin{align}
		U''+\bigg[E^2-M^2&+M\omega(1+2sj)-M^2\omega^2r^2
		\notag\\
		&{}-\frac{j(j-s)}{r^2}\bigg]U=0.
		\label{eq:flat-radial}
	\end{align}
	With the substitution $x=M\omega r^2$ and $U=r^{\abs{m_{js}}+1/2}\ee^{-x/2}w(x)$, Eq.~\eqref{eq:flat-radial} reduces to the associated Laguerre equation $x\,w''+(\abs{m_{js}}+1-x)w'+n\,w=0$, so the regular modes are
	\begin{equation}
		U_{njs}\propto r^{\abs{m_{js}}+1/2}
		\ee^{-M\omega r^2/2}
		L_n^{\abs{m_{js}}}(M\omega r^2),
		\label{eq:flat-mode}
	\end{equation}
	with the quantization condition $E^2-M^2=2M\omega(2n+\abs{m_{js}}-s m_{js})$, i.e.
	\begin{equation}
		E^2=M^2+2M\omega\left(2n+\abs{m_{js}}-sm_{js}\right),
		\qquad n=0,1,2,\dots
		\label{eq:flat-spectrum}
	\end{equation}
	For the oscillator sign adopted here, the channel with $2n+\abs{m_{js}}-sm_{js}=0$ gives $E=+M$, a normalizable isolated threshold state; the formal $E=-M$ partner does not belong to the same normalizable domain and must not be counted as a physical level.
	
	\subsection{Ideal cone}
	\label{sec:cone}
	
	For $N=a$ and $L=br$ the lapse is constant, so $\epsilon_\pm=E/a\pm M$, and the effective angular index is rescaled by the slope.  Introducing
	\begin{equation}
		\lambda_{js}=\frac{j}{b}-\frac{s}{2},
		\label{eq:lambda}
	\end{equation}
	the same Laguerre reduction as in Sec.~\ref{sec:flat} carries through with $\abs{m_{js}}\to\abs{\lambda_{js}}$ and $E^2\to E^2/a^2$, giving the exact spectrum
	\begin{equation}
		\frac{E^2}{a^2}=M^2+2M\omega
		\left(2n+\abs{\lambda_{js}}-s\lambda_{js}\right),
		\label{eq:cone-spectrum}
	\end{equation}
	with $U\propto r^{\abs{\lambda_{js}}+1/2}\ee^{-M\omega r^2/2}L_n^{\abs{\lambda_{js}}}(M\omega r^2)$.  The constant $a$ rescales the coordinate energy, whereas $b$ changes the effective angular index through Eq.~\eqref{eq:lambda}.  For a resolved string, $L(r)$ approaches $br+c$ only asymptotically, while both $L(r)$ and $N(r)$ vary through the core.  This variable-coefficient structure destroys the exact Laguerre reduction; the finite-core information is carried by the complete profiles rather than by the asymptotic offset $c$ alone.  On the singular cone we choose the regular branch, retaining the larger Frobenius power in Eq.~\eqref{eq:frobenius}.  This is the branch continuously selected when a smooth resolved core is shrunk toward the conical limit.
	
	\section{Numerical formulation}
	\label{sec:numerical}
	
	\subsection{Laguerre--Galerkin discretization}
	\label{sec:galerkin}
	
	We expand the two components in separate normalized radial oscillator bases that build in the correct axis power and Gaussian tail,
	\begin{equation}
		\phi_n^{(\alpha)}(r)=
		\left[\frac{2\kappa^{\alpha+1}n!}{\Gamma(n+\alpha+1)}\right]^{1/2}
		r^{\alpha+1/2}\ee^{-\kappa r^2/2}
		L_n^{\alpha}(\kappa r^2),
		\label{eq:basis}
	\end{equation}
	with $\kappa=M\omega$, and $\alpha=\alpha_U$ for $U$ and $\alpha=\alpha_V$ for $V$ from Eq.~\eqref{eq:alphas}.  Because the exponent $\alpha$ is computed from the measured axis slope $\ell$, the basis automatically carries the correct conical or resolved-core power and never evaluates $1/L$ at $r=0$.  Projecting the operators of Eq.~\eqref{eq:HW} onto this basis, and writing the matrix elements
	\begin{align}
		(\mathsf S_U)_{ij}&=\int\dd r\,\phi_i^{(\alpha_U)}\phi_j^{(\alpha_U)},
		\label{eq:matelemS}\\
		(\mathsf W_U)_{ij}&=\int\frac{\dd r}{N}\,\phi_i^{(\alpha_U)}\phi_j^{(\alpha_U)},
		\label{eq:matelemW}\\
		\mathsf C_{ij}&=\int\dd r\,\phi_i^{(\alpha_U)}\left(-\partial_r+B\right)\phi_j^{(\alpha_V)},
		\label{eq:matelem}
	\end{align}
	(with $\mathsf S_V$ and $\mathsf W_V$ defined analogously) the eigenproblem~\eqref{eq:gevp} becomes the algebraic pencil
	\begin{equation}
		\bm H\bm c=E\,\bm W\bm c,
		\label{eq:matrix-gevp}
	\end{equation}
	with the Hermitian and weight blocks
	\begin{equation}
		\bm H=\begin{pmatrix}M\mathsf S_U&\mathsf C\\ \mathsf C^{\!\top}&-M\mathsf S_V\end{pmatrix},
		\qquad
		\bm W=\begin{pmatrix}\mathsf W_U&0\\0&\mathsf W_V\end{pmatrix}.
		\label{eq:matrix-blocks}
	\end{equation}
	The off-diagonal blocks are constructed as transposes of one another, so $\bm H$ is symmetric by construction, and $\bm W$ is symmetric positive definite because $N>0$.  We solve the pencil by the congruence transformation $\bm A=\bm W^{-1/2}\bm H\bm W^{-1/2}$, which is symmetric and shares the spectrum of the pencil; its eigenvectors $\bm u$ return the physical coefficients through $\bm c=\bm W^{-1/2}\bm u$, normalized so that $\bm c^{\!\top}\bm W\bm c=1$ in accordance with Eq.~\eqref{eq:norm}.  Keeping $\bm W$ positive definite is what protects this step, and its smallest eigenvalue is monitored throughout the calculation.
	
	\subsection{Pruefer-angle shooting}
	\label{sec:pruefer}
	
	For an independent calculation we track a phase rather than an amplitude.  Writing
	\begin{equation}
		U=R(r)\cos\theta(r),
		\qquad
		V=R(r)\sin\theta(r),
		\label{eq:pruefer}
	\end{equation}
	and inserting into Eqs.~\eqref{eq:UV1}--\eqref{eq:UV2}, the amplitude $R$ drops out of the ratio $V/U=\tan\theta$.  Using $\theta'=(V'U-U'V)/R^2$ together with $U'=-BU+\epsilon_+V$ and $V'=BV-\epsilon_-U$ gives $V'U-U'V=2BUV-\epsilon_-U^2-\epsilon_+V^2$, so
	\begin{equation}
		\theta'=B\sin(2\theta)
		-\epsilon_-\cos^2\theta
		-\epsilon_+\sin^2\theta.
		\label{eq:theta}
	\end{equation}
	We integrate the regular solution from $r_{\min}$ to a matching point $r_m$, using the Frobenius power to set the near-axis phase $\theta(r_{\min})=\arctan(V/U)$, and the decaying solution inward from $r_{\max}$ in the reversed coordinate $x=r_{\max}-r$.  A bound state exists when the two phases meet modulo $\pi$, i.e. when
	\begin{equation}
		\Delta(E)=\sin\!\left[\theta_{\mathrm L}(r_m)-\theta_{\mathrm R}(r_m)\right]=0.
		\label{eq:mismatch}
	\end{equation}
	A coarse energy scan brackets the roots and bisection refines them.  Guard digits are used in the axis initialization because the numerator of $V/U$ contains a leading-order cancellation between the $M\omega r$ and $sj/L$ pieces of $B$.  Since only the phase is propagated, the method is immune to the exponentially growing amplitude that would otherwise contaminate a direct forward integration.
	
	\subsection{Direct matching of the second-order equations}
	\label{sec:secondorder}

	As a third and algebraically independent implementation, we integrate Eqs.~\eqref{eq:upper-decoupled} and \eqref{eq:lower-decoupled} directly.  For the upper component, a regular solution $(U_{\rm L},U'_{\rm L})$ is propagated outward from the Frobenius expansion, and a Gaussian-decaying solution $(U_{\rm R},U'_{\rm R})$ is propagated inward from the outer boundary.  The scale-independent matching function is the normalized Wronskian
	\begin{equation}
		\mathcal M_U(E)=
		\frac{U_{\rm L}U'_{\rm R}-U'_{\rm L}U_{\rm R}}
		{\sqrt{U_{\rm L}^2+U_{\rm L}'^{\,2}}\,
		 \sqrt{U_{\rm R}^2+U_{\rm R}'^{\,2}}},
		\label{eq:MU}
	\end{equation}
	evaluated at the same matching radius used by the Pruefer calculation.  Its zeros determine the spectrum.  An analogous function $\mathcal M_V(E)$ is constructed from Eq.~\eqref{eq:lower-decoupled}.  The latter is used only when $\epsilon_-(r;E)$ does not vanish on the integration interval; the threshold channel is instead handled by the regular first-order system.  Agreement among the generalized eigenproblem, the Pruefer zeros, and $\mathcal M_U(E)=0$, together with the independent $\mathcal M_V$ benchmark, tests both spinor components and the reconstruction relations.

Candidate Galerkin levels are also tested for radial localization. A boundary-localized pseudo-state can occur close to the formal negative threshold $E=-aM$; it is removed by a tail-leakage criterion. The diagnostic and representative densities are given in the Supplemental Material.

\section{Numerical Analysis}
\label{sec:results}

We use $M=1$, $j=1/2$, and $s=+1$ unless stated otherwise. For the exact benchmark we set $\omega=1/4$; the ideal-cone case has $a=0.96$ and $b=0.72$. The first three positive levels obtained by two-sided Pruefer shooting are compared with the analytic Minkowski and ideal-cone spectra in Table~\ref{tab:benchmarks}. The errors range from $10^{-11}$ to a few $10^{-9}$. For this particular channel, $\lambda_{js}=j/b-s/2>0$, so $|\lambda_{js}|-s\lambda_{js}=0$ and the exact energies are independent of $b$; this benchmark therefore tests the lapse rescaling and the regular/conical radial implementation, not the spectral dependence on the deficit. An independent Laguerre--Galerkin calculation reproduces the analytic positive and negative branches with maximum relative deviations $4.04\times10^{-5}$ in Minkowski space and $1.01\times10^{-5}$ on the cone. These basis-truncation errors are quoted only as numerical diagnostics and are not assigned physical significance.

\begin{table}[tbp]
    \caption{Exact and Pruefer-shooting energies for the first three positive levels.}
    \label{tab:benchmarks}
    \begin{ruledtabular}
        \begin{tabular}{cccc}
            background & level & $E_{\rm exact}$ & $E_{\rm shoot}$\\
            \hline
            flat & 0 & $1.0000000000$ & $1.0000000000$\\
            flat & 1 & $1.4142135624$ & $1.4142135622$\\
            flat & 2 & $1.7320508076$ & $1.7320508104$\\
            cone & 0 & $0.9600000000$ & $0.9599999999$\\
            cone & 1 & $1.3576450199$ & $1.3576450207$\\
            cone & 2 & $1.6627687753$ & $1.6627687746$\\
        \end{tabular}
    \end{ruledtabular}
\end{table}

Direct matching of the decoupled second-order equations gives the same benchmark energies to better than $4\times10^{-10}$. The Supplemental Material reports further tests on an exactly regular variable-coefficient geometry, basis-size convergence, localization of the threshold pseudo-state, and cross-validation on numerical vortex profiles.

\section{Self-gravitating vortex background}
	\label{sec:vortex}

	\subsection{Einstein--Abelian-Higgs system}
	The physical background is generated by a Nielsen--Olesen vortex minimally coupled to gravity~\cite{NPB.1973.61.45,PRD.1985.32.1323,PRD.1999.60.125012}.  With $\Phi=\eta X(r)\ee^{\ii \nu\varphi}$ and $A_\varphi=(\nu/e)[1-P(r)]$, where $\nu\in\mathbb Z$ is the vortex winding number, the matter Lagrangian is
	\begin{equation}
		\Lag_{\rm m}=\abs{D_\mu\Phi}^2-\frac{1}{4}F_{\mu\nu}F^{\mu\nu}
		-\frac{\lambda}{4}\left(\abs\Phi^2-\eta^2\right)^2.
		\label{eq:AHlag}
	\end{equation}
	With canonical fluctuations about the vacuum,
	\begin{equation}
		m_{\rm s}^2=\lambda\eta^2,
		\qquad m_{\rm v}^2=2e^2\eta^2,
		\qquad
		\beta\equiv\frac{\lambda}{e^2}=2\frac{m_{\rm s}^2}{m_{\rm v}^2}.
		\label{eq:beta-correct}
	\end{equation}
	The radial variables used in Eqs.~\eqref{eq:vX}--\eqref{eq:constraint} are made dimensionless with $e\eta$, so that $r=e\eta\,r_{\rm phys}$ and $L=e\eta\,L_{\rm phys}$.  The Dirac parameters on these backgrounds are scaled consistently as $M=M_{\rm phys}/(e\eta)$ and $\omega=\omega_{\rm phys}/(e\eta)$.  Thus, in the units $e=\eta=1$ used below, the scalar and vector length scales expressed in this dimensionless coordinate are $\xi_{\rm s}=1/\sqrt\beta$ and $\xi_{\rm v}=1/\sqrt2$, respectively.  Because neither length alone uniquely defines the nonlinear vortex width, comparisons between cores use the directly measured Higgs half-amplitude radius
	\begin{equation}
		X(r_{1/2})=\frac12.
		\label{eq:rhalf}
	\end{equation}
	The gravitational coupling is $\varepsilon=8\pi G\eta^2$.  Computing the Einstein and matter tensors gives
	\begin{align}
		X''+\Big(\tfrac{2N'}{N}+\tfrac{L'}{L}\Big)X'&=\tfrac{\beta}{2}X(X^2-1)+\frac{\nu^2P^2X}{L^2},
		\label{eq:vX}\\
		P''+\Big(\tfrac{2N'}{N}-\tfrac{L'}{L}\Big)P'&=2X^2P,
		\label{eq:vP}\\
		N''&=-\tfrac{\varepsilon}{2}N T^{\varphi}{}_{\varphi}-\frac{N'^2}{2N},
		\label{eq:vN}\\
		L''&=-\varepsilon L\rho-\frac{LN''}{N}-\frac{N'L'}{N},
		\label{eq:vL}
	\end{align}
	where
	\begin{align}
		\rho&=X'^2+\frac{\nu^2P'^2}{2L^2}+\frac{\nu^2X^2P^2}{L^2}
		+\tfrac{\beta}{4}(X^2-1)^2,
		\label{eq:rho}\\
		T^{\varphi}{}_{\varphi}&=X'^2-\frac{\nu^2P'^2}{2L^2}-\frac{\nu^2X^2P^2}{L^2}
		+\tfrac{\beta}{4}(X^2-1)^2.
		\label{eq:Tphi}
	\end{align}
	The remaining Einstein equation,
	\begin{align}
		\frac{N'^2}{N^2}+\frac{2N'L'}{NL}&=-\varepsilon T^r{}_r,
		\notag\\
		T^r{}_r&=-X'^2-\frac{\nu^2P'^2}{2L^2}
		+\frac{\nu^2X^2P^2}{L^2}
		\notag\\
		&\quad+\tfrac{\beta}{4}(X^2-1)^2.
		\label{eq:constraint}
	\end{align}
	is not used as an evolution equation and provides an independent constraint check.

	The fermion couples to the vortex only through the metric.  It has neither a minimal coupling to $A_\mu$ nor a Yukawa coupling to $\Phi$, and its stress tensor is neglected.  The calculation is therefore a test-field bound-state problem on a self-consistent gravitating-vortex background, not a backreacting Einstein--Dirac--Higgs solution.

	\subsection{Regular boundary-value formulation}
	To remove the coordinate singularities at the axis, the numerical solver uses
	\begin{equation}
		X=r x(r),\qquad P=1-r^2q(r),\qquad L=r\ell(r).
		\label{eq:regular-vars}
	\end{equation}
	The regular conditions are $x'(0)=q'(0)=N'(0)=\ell'(0)=0$, $N(0)=\ell(0)=1$, together with $X\to1$ and $P\to0$ at the outer boundary.  This formulation enforces $L'(0)=1$ without evaluating terms of the form $1/L$ at the axis.

The regular formulation is solved as a boundary-value problem and the unused Einstein equation~\eqref{eq:constraint} is monitored as a constraint. For the backgrounds used in the fixed-cone scan its bulk residual remains below $8\times10^{-6}$, with root-mean-square residuals of order $10^{-6}$ or smaller outside the axis initialization region. Additional background-solver checks are summarized in the Supplemental Material.

\begin{figure}[tbp]
		\includegraphics[width=\columnwidth]{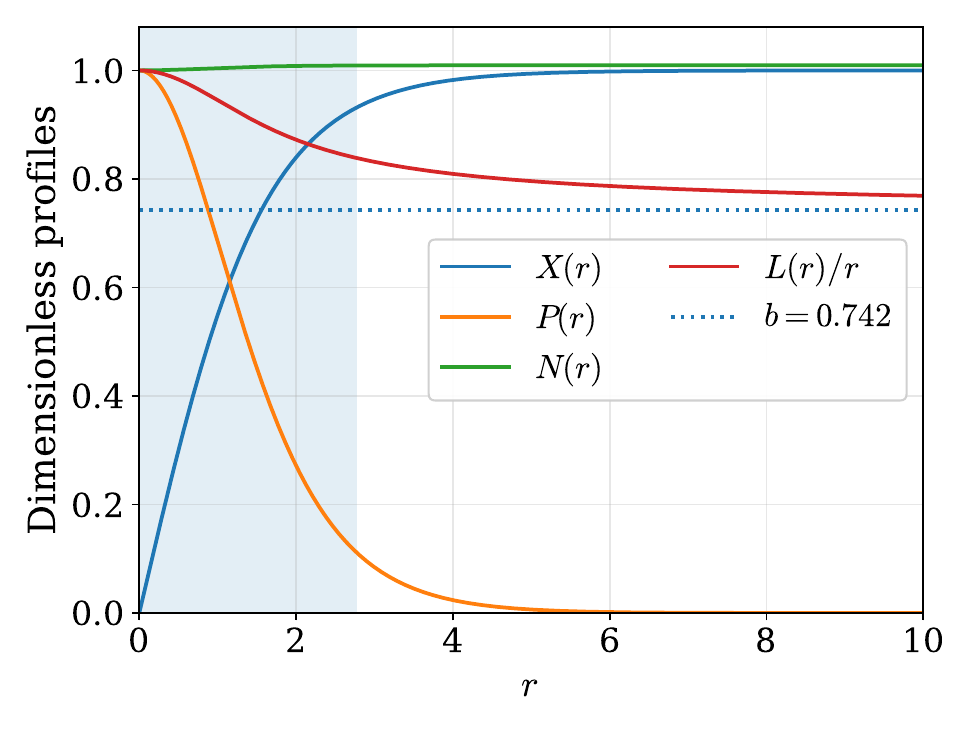}
		\caption{Self-gravitating Nielsen--Olesen vortex for $\beta=1$, $\varepsilon=0.3$, and $\nu=1$.  The Higgs profile rises, the gauge profile falls, and $L(r)/r$ relaxes from unity at the regular axis to the asymptotic slope $b=0.742$.  The shaded region indicates twice the larger of the Higgs and gauge half-amplitude radii and serves only as a visual guide to the transition region; the quantitative core measure used in the spectral scans is the Higgs half-amplitude radius $r_{1/2}$.}
		\label{fig:vortex}
	\end{figure}

	The representative solution in Fig.~\ref{fig:vortex} has $a=1.0093$, $b=0.7421$, and $r_{1/2}=0.889$. The spectral calculations use the one-dimensional profiles $N(r)$ and $L(r)$ shown together with the vortex matter profiles in that figure.

\section{Confinement spectra in the physical vortex}
	\label{sec:spectra}

For each numerical vortex, let $a=N(\infty)$ and define the asymptotically normalized lapse and energy by
	\begin{equation}
		\bar N(r)=\frac{N(r)}{a},
		\qquad
		\mathcal E=\frac{E}{a}.
		\label{eq:normalized-energy}
	\end{equation}
	This is the constant time rescaling that sets $\bar N(\infty)=1$.  The ideal comparison cone is therefore characterized by $(\bar a,b)=(1,b)$.  Relative shifts are unchanged by the rescaling when the resolved core and its cone share the same $a$, while Eq.~\eqref{eq:normalized-energy} is essential when different vortex families are compared at fixed $b$.

On representative numerical Einstein--Abelian-Higgs backgrounds, Galerkin, Pruefer, and direct second-order determinations of the normalized energies agree to $1.1\times10^{-8}$ or better; the level-by-level comparison is given in the Supplemental Material.

\subsection{Reference vortex and confinement scale}
\begin{figure}[tbp]
		\includegraphics[width=\columnwidth]{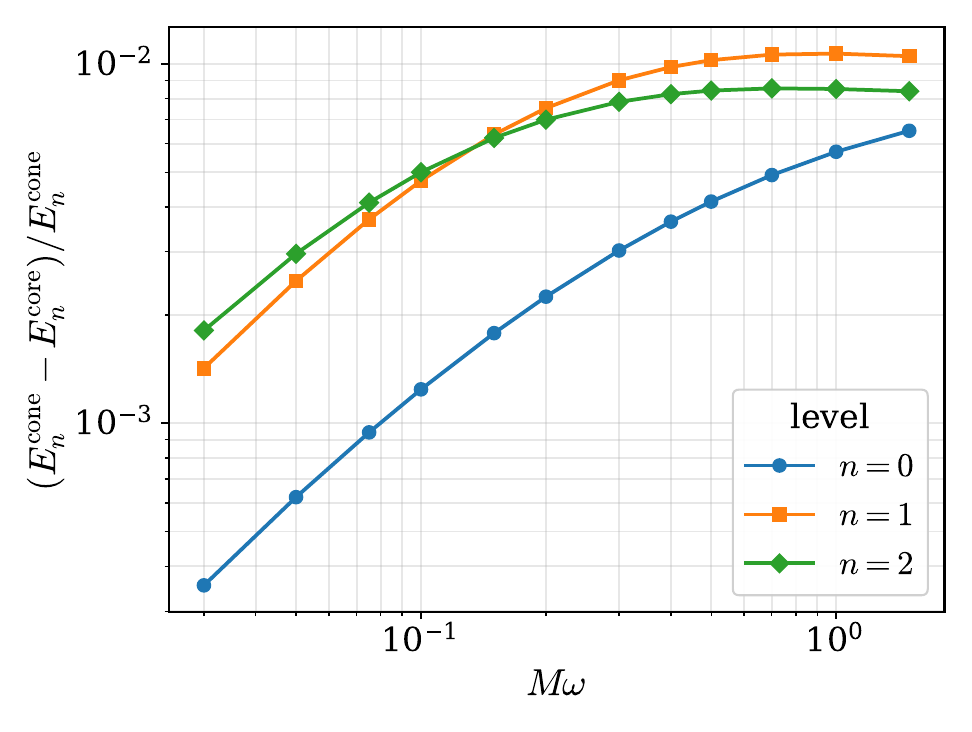}
		\caption{Finite-core shift versus confinement scale for the reference vortex $\beta=1$, $\varepsilon=0.3$.  Positive values mean that the resolved-core level lies below the ideal-cone level with the same asymptotic lapse $a$ and angular slope $b$.}
		\label{fig:cross}
	\end{figure}

	For the reference family $\beta=1$, $\varepsilon=0.3$, Fig.~\ref{fig:cross} shows
	$(E_n^{\rm cone}-E_n^{\rm core})/E_n^{\rm cone}$ as $M\omega$ is varied.  The three displayed resolved-core levels are below their cone counterparts.  The correction tends to zero in the weak-confinement regime and grows through the core-sensitive region.  At the largest frequencies the excited branches approach a percent-level plateau and show mild nonmonotonic saturation rather than an indefinitely increasing shift.

\subsection{Core dependence at a fixed normalized cone}
	We tune $\varepsilon(\beta)$ to the target slope $b=0.82$ to the quoted numerical accuracy, normalize the asymptotic lapse according to Eq.~\eqref{eq:normalized-energy}, and measure the core by $r_{1/2}$.  Table~\ref{tab:fixedcone} shows that the raw lapse $a$ varies slightly across the family, which is why comparing $\mathcal E=E/a$ is required.  The last-digit deviation $b=0.81999$ in one entry reflects the background-solver tolerance; after lapse normalization the cases represent the same target asymptotic cone $(\bar a,b)=(1,0.82)$ within that accuracy.

	\begin{table}[tbp]
		\caption{Vortex parameters in the fixed-cone scan, tuned to the target slope $b=0.82$ within the numerical tolerance of the background solver.}
		\label{tab:fixedcone}
		\begin{ruledtabular}
			\begin{tabular}{ccccc}
				$\beta$ & $\varepsilon$ & $a$ & $b$ & $r_{1/2}$\\
				\hline
				$0.5$ & $0.24079$ & $1.01230$ & $0.82000$ & $1.09791$\\
				$1.0$ & $0.20878$ & $1.00625$ & $0.82000$ & $0.86838$\\
				$2.0$ & $0.18000$ & $1.00000$ & $0.82000$ & $0.67940$\\
				$4.0$ & $0.15456$ & $0.99363$ & $0.81999$ & $0.52490$\\
			\end{tabular}
		\end{ruledtabular}
	\end{table}

\begin{figure}[tbp]
		\includegraphics[scale=0.46]{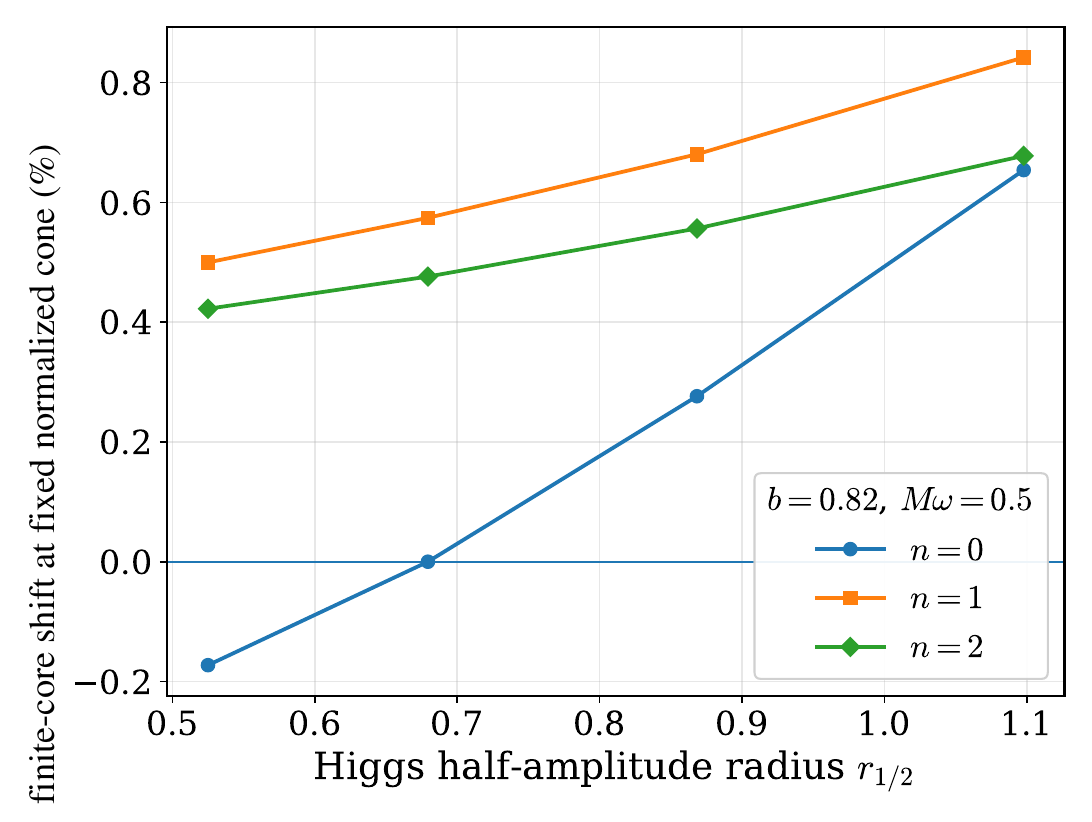}
		\caption{Finite-core shift (in percent) at the target normalized cone $(\bar a,b)=(1,0.82)$, realized to the quoted numerical accuracy, and $M\omega=0.5$, plotted against the measured Higgs half-amplitude radius.  The four sampled backgrounds bracket a ground-state sign change but do not determine the crossing radius more finely. Lines connecting the sampled points are guides to the eye.}
		\label{fig:core}
	\end{figure}

	Figure~\ref{fig:core} shows that the normalized levels still depend on the internal profile.  Across the four sampled backgrounds, as $r_{1/2}$ decreases from $1.098$ to $0.525$, the ground-state shift changes from $+0.654\%$ to $-0.173\%$.  One sampled point lies close to zero, and the positive and negative values bracket a sign change within this fixed-cone family; the four-point scan is not used to assign a more precise crossing radius.  The first two excited levels remain below the cone and vary by several tenths of a percent.  Because $\bar a=1$ exactly after normalization and $b$ is fixed to the quoted numerical accuracy, this variation is a finite-core effect rather than a lapse-normalization or deficit-angle artifact.

A complementary scan along the reference $\beta=1$ family, where the asymptotic deficit and the internal core vary together, is documented in the Supplemental Material.  The same supplement also shows that nonzero finite-core corrections persist in additional $(j,s)$ channels of the reference vortex.

The mechanism is visible directly in the radial operator.  The normalized lapse controls the local energy and the weight, while $L(r)$ controls the spin--angular term.  Near a resolved axis $L/r\to1$, whereas the singular comparison cone has $L/r=b$.  A strongly localized wave function therefore samples a centrifugal and redshift profile that cannot be reconstructed from the asymptotic constants alone.

	\section{Conclusions}
\label{sec:conclusions}

We have formulated the transverse Dirac oscillator as a test field on a self-consistent gravitating Abelian-Higgs vortex and written the radial equations as a generalized Hermitian eigenvalue problem with positive weight. The numerical spectrum reproduces the exact Minkowski and ideal-cone limits before the method is applied to resolved vortex geometries.

For the reference vortex, the finite-core correction grows toward the percent level when the confinement length becomes comparable to the core scale. The sharper test is provided by the fixed-cone family: after normalizing the asymptotic lapse and holding $b$ fixed to numerical accuracy, the sampled backgrounds retain distinct spectra and bracket a sign change of the ground-state correction. The confined spectrum therefore depends on the resolved profiles $N(r)$ and $L(r)$, not only on the far-field cone.

The present calculation neglects fermionic backreaction and direct gauge or Yukawa couplings. Non-Abelian vortices and nonzero longitudinal momentum are further natural extensions.

\section*{Data availability}
	The numerical values used in the tabulated checks are reported in the manuscript and Supplemental Material.

	\begin{acknowledgments}
	The author acknowledges support from Conselho Nacional de Desenvolvimento Cient\'{i}fico e Tecnol\'{o}gico (CNPq) (grant 305427/2026-1), Funda\c c\~ao de Amparo \`a Pesquisa e ao Desenvolvimento Cient\'{i}fico e Tecnol\'{o}gico do Maranh\~ao (FAPEMA) (grant UNIVERSAL-06395/22), and Coordena\c c\~ao de Aperfei\c coamento de Pessoal de N\'{i}vel Superior (CAPES) - Brazil (Finance Code 001).
	\end{acknowledgments}

%

\end{document}


\begin{center}
{\large\bfseries Supplemental Material for ``Dirac oscillator in a self-gravitating cosmic string: confinement spectra beyond the conical approximation''}\\[0.6em]
Edilberto O. Silva
\end{center}
\vspace{0.8em}

\section*{I. Extended numerical validation}

The main text retains one exact spectral benchmark and the physical figures needed for the argument. The additional tests collected here document the numerical stability of the calculation without interrupting the main physical discussion.

Unless stated otherwise, the spectral tests in Secs.~I.A--I.C use $M=1$, $\omega=1/4$, $j=1/2$, and $s=+1$.  Section~I.D and the additional physical scans state their values of $M\omega$ explicitly.

\subsection*{A. Direct second-order benchmark}

The decoupled second-order equations provide an implementation independent of both the generalized Laguerre--Galerkin eigenproblem and the first-order Pruefer shooting calculation.  The conical case has $a=0.96$ and $b=0.72$.  In this channel $\lambda_{js}=j/b-s/2>0$, so $|\lambda_{js}|-s\lambda_{js}=0$ and the analytic energies are independent of $b$.  The test therefore probes the lapse rescaling and the conical radial implementation rather than the spectral dependence on the deficit. Table~\ref{tab:s-secondorder} compares the upper-component roots with the exact flat and conical energies. For excited states, the lower-component matching residual is also evaluated at the exact energy. The ground-state entries are omitted from the latter test because they belong to the threshold channel in which division by the corresponding energy factor is not admissible.

\begin{table}[htbp]
\caption{Direct second-order checks. The third column is the absolute error of the upper-component root, and the fourth is the absolute lower-component matching residual evaluated at the exact energy.}
\label{tab:s-secondorder}
\begin{ruledtabular}
\begin{tabular}{cccc}
background & $n$ & $\abs{E_U^{(2)}-E_{\rm ex}}$ & $\abs{\mathcal M_V(E_{\rm ex})}$\\
\hline
flat & 0 & $3.97\times10^{-10}$ & --\\
flat & 1 & $2.52\times10^{-10}$ & $5.81\times10^{-13}$\\
flat & 2 & $2.06\times10^{-10}$ & $2.13\times10^{-13}$\\
cone & 0 & $1.55\times10^{-11}$ & --\\
cone & 1 & $1.16\times10^{-11}$ & $1.18\times10^{-13}$\\
cone & 2 & $1.03\times10^{-11}$ & $5.78\times10^{-14}$\\
\end{tabular}
\end{ruledtabular}
\end{table}
\FloatBarrier

\subsection*{B. Exactly regular variable-coefficient benchmark}

A regular finite-core test geometry is defined by
\begin{align}
N(r)&=1+(a-1)\frac{r^2}{r^2+r_c^2},\\
L(r)&=r+(b-1)\frac{r^3}{r^2+r_c^2}+c\tanh^3\!\left(\frac{r}{r_c}\right),
\end{align}
with
\begin{equation}
a=0.94,\qquad b=0.72,\qquad c=0.35,\qquad r_c=1.2.
\end{equation}
These functions satisfy $N'(0)=0$, $L(0)=0$, and $L'(0)=1$ and approach $N\to a$ and $L\to br+c$. Their explicit plots are not included because those properties follow directly from the analytic definitions; the nontrivial test is the agreement of independent spectral formulations.

\begin{table}[htbp]
\caption{Lowest positive levels in the regular variable-coefficient test geometry.}
\label{tab:s-toy}
\begin{ruledtabular}
\begin{tabular}{ccccc}
$n$ & $E_{\rm Gal}$ & $E_{\rm Pr}$ & $E_U^{(2)}$ & $\abs{E_{\rm Pr}-E_U^{(2)}}$\\
\hline
0 & $0.9631329282$ & $0.9631329210$ & $0.9631329207$ & $3.17\times10^{-10}$\\
1 & $1.3400233986$ & $1.3400233680$ & $1.3400233677$ & $2.40\times10^{-10}$\\
2 & $1.6366462030$ & $1.6366461044$ & $1.6366461042$ & $1.96\times10^{-10}$\\
\end{tabular}
\end{ruledtabular}
\end{table}
\FloatBarrier

\begin{figure}[htbp]
\centering
\includegraphics[width=0.68\textwidth]{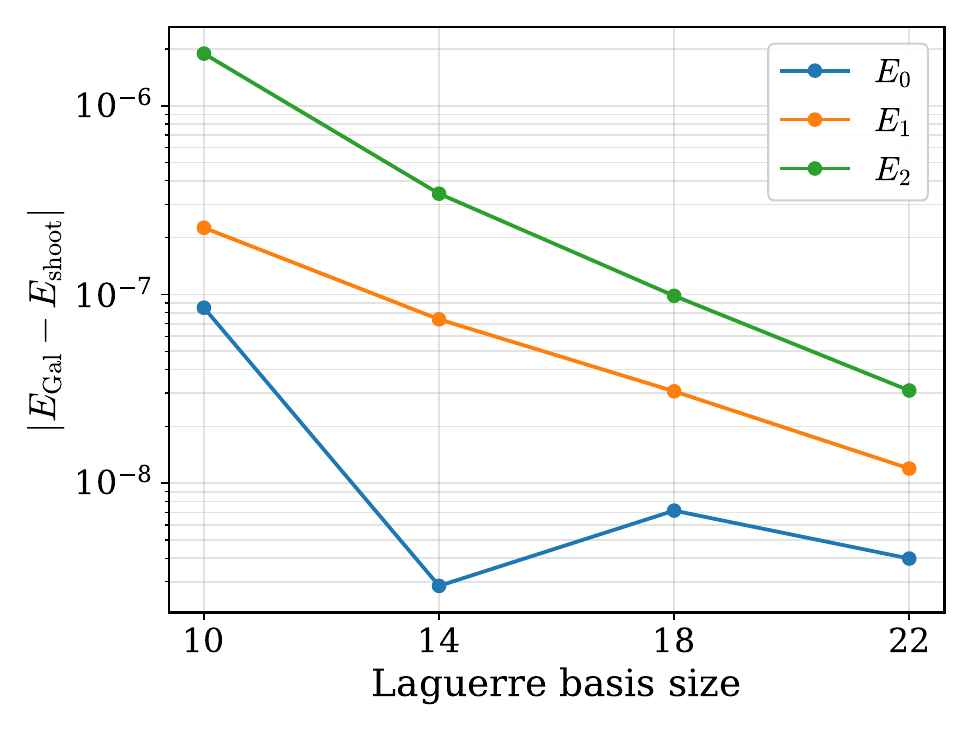}
\caption{Absolute difference between Galerkin and Pruefer energies for the three lowest positive levels of the regular variable-coefficient test geometry as the Laguerre basis is enlarged.}
\label{fig:s-convergence}
\end{figure}

Figure~\ref{fig:s-convergence} records the basis-size test. The $n=0$ difference is not strictly monotonic, but all three levels stabilize at a scale far below the finite-core spectral shifts discussed in the main text.

\FloatBarrier

\subsection*{C. Localization and threshold pseudo-state}

Finite-basis diagonalization can return a boundary-localized pseudo-state close to the formal negative threshold. Candidate levels are therefore tested with the tail probability
\begin{equation}
P_{\rm tail}=\frac{\displaystyle\int_{r_t}^{r_{\max}}\frac{dr}{N(r)}\left(|U|^2+|V|^2\right)}{\displaystyle\int_{0}^{r_{\max}}\frac{dr}{N(r)}\left(|U|^2+|V|^2\right)},
\end{equation}
where $r_t$ is placed in the last $10\%$ of the radial interval, together with the density ratio at the outer boundary. A state is retained only when both leakage measures are below $10^{-3}$.

\begin{figure}[htbp]
\centering
\includegraphics[width=0.68\textwidth]{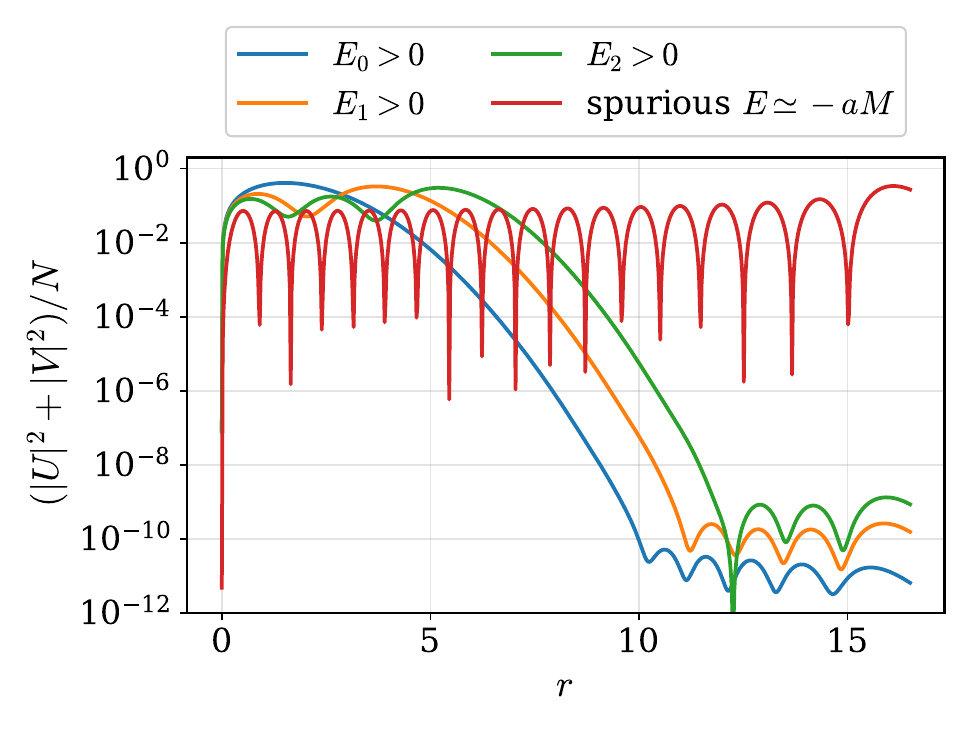}
\caption{Weighted radial densities in the regular variable-coefficient geometry. The positive levels are localized, whereas the formal negative-threshold partner is concentrated near the numerical boundary and is rejected by the leakage criterion.}
\label{fig:s-densities}
\end{figure}

In the regular test geometry the positive ground state has $P_{\rm tail}=2.06\times10^{-11}$. The formal negative-threshold partner near $E=-0.943904$ instead has mean radius $\langle r\rangle=10.88$, peaks near $r=16.11$, has $P_{\rm tail}=0.317$, and a boundary-density ratio $0.800$. Figure~\ref{fig:s-densities} makes the distinction explicit.

\FloatBarrier

\subsection*{D. Cross-validation on physical vortex profiles}

The three spectral implementations were also compared directly on numerical Einstein--Abelian-Higgs backgrounds. Table~\ref{tab:s-physical-cross} shows representative normalized energies for backgrounds tuned to the target $b=0.82$ (within the numerical tolerance of the background solver) at $M\omega=0.5$. The maximum Galerkin--Pruefer difference is $1.03\times10^{-8}$ and the maximum Pruefer--second-order difference is $1.01\times10^{-9}$.

\begin{table}[htbp]
\caption{Cross-validation of asymptotically normalized energies on physical vortex profiles.}
\label{tab:s-physical-cross}
\begin{ruledtabular}
\begin{tabular}{ccccc}
$\beta$ & $n$ & $\mathcal E_{\rm Pr}$ & $\abs{\mathcal E_{\rm Gal}-\mathcal E_{\rm Pr}}$ & $\abs{\mathcal E_{\rm Pr}-\mathcal E_U^{(2)}}$\\
\hline
$0.5$ & 0 & $1.0056760599$ & $2.5\times10^{-9}$ & $8.0\times10^{-10}$\\
$0.5$ & 1 & $1.7385821695$ & $7.3\times10^{-9}$ & $4.0\times10^{-10}$\\
$0.5$ & 2 & $2.2482197417$ & $1.03\times10^{-8}$ & $3.0\times10^{-10}$\\
$1.0$ & 0 & $1.0034739546$ & $2.5\times10^{-9}$ & $1.0\times10^{-9}$\\
$1.0$ & 1 & $1.7310254370$ & $7.2\times10^{-9}$ & $5.0\times10^{-10}$\\
$1.0$ & 2 & $2.2375335146$ & $1.01\times10^{-8}$ & $4.0\times10^{-10}$\\
\end{tabular}
\end{ruledtabular}
\end{table}
\FloatBarrier

The quoted levels are stable against enlargement of the radial domain and Laguerre basis and against tighter quadrature and integration tolerances. The generalized weight matrix remains positive definite and the algebraic Hermiticity residual is at floating-point level.

\section*{II. Background-solver checks}

The regular variables $X=rx(r)$, $P=1-r^2q(r)$, and $L=r\ell(r)$ remove the coordinate singularities at the axis. The imposed boundary values reproduce one unit of magnetic flux, and the limit of vanishing gravitational coupling recovers the flat Nielsen--Olesen system. In a weak-field test at $\varepsilon=0.02$, the asymptotic-slope deficit agrees with the leading energy-density estimate to about $0.4\%$. For the backgrounds used in the fixed-cone scan, the unused Einstein constraint has a bulk residual below $8\times10^{-6}$ and a root-mean-square residual of order $10^{-6}$ or smaller outside the axis initialization region.

\section*{III. Additional physical scans}

\subsection*{A. Correlated deficit/core family}

\begin{figure}[htbp]
\centering
\includegraphics[width=0.68\textwidth]{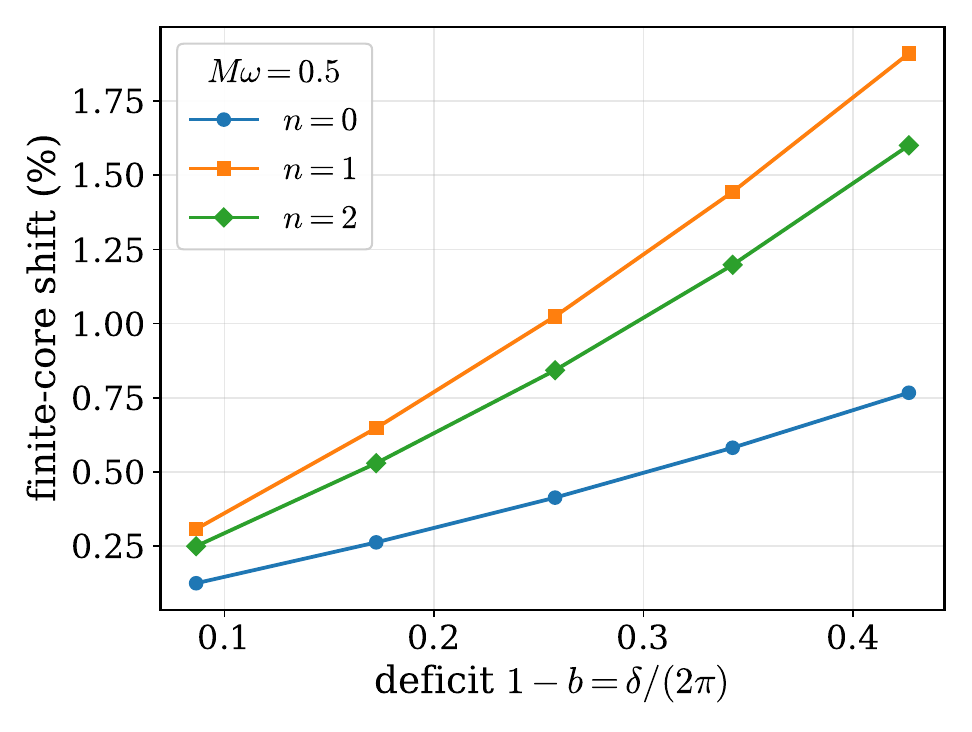}
\caption{Finite-core shift along the $\beta=1$ vortex family at $M\omega=0.5$, plotted against $1-b=\delta/(2\pi)$. The scan correlates changes of the asymptotic deficit with changes of the internal profiles.}
\label{fig:s-deficit}
\end{figure}

At fixed $M\omega=0.5$ and $\beta=1$, varying the gravitational coupling changes both the asymptotic deficit and the internal vortex profiles. Figure~\ref{fig:s-deficit} therefore represents a correlated family, not an isolated dependence on $b$. Along this family the finite-core correction grows in magnitude; the first excited level reaches about $1.9\%$ at the largest displayed value of $1-b$. The fixed-cone family in the main text is the cleaner test because it holds the asymptotic cone fixed while varying the resolved core.

\FloatBarrier

\subsection*{B. Additional angular channels}

Nonzero finite-core corrections are also present outside the channel used in the main scans. Table~\ref{tab:s-channels} gives the percentage shifts for two additional $(j,s)$ sectors of the reference vortex at $M\omega=0.5$.  Because these entries refer to a single background rather than to a scan over core radii, they establish persistence across angular channels but are not used to infer the same core-size dependence in each sector.

\begin{table}[htbp]
\caption{Finite-core shifts in percent for the reference vortex at $M\omega=0.5$.}
\label{tab:s-channels}
\begin{ruledtabular}
\begin{tabular}{cccc}
$(j,s)$ & $n=0$ & $n=1$ & $n=2$\\
\hline
$(1/2,+1)$ & $0.414$ & $1.025$ & $0.843$\\
$(3/2,+1)$ & $0.180$ & $1.647$ & $1.559$\\
$(1/2,-1)$ & $4.293$ & $2.541$ & $1.800$\\
\end{tabular}
\end{ruledtabular}
\end{table}
\FloatBarrier